\documentclass[prd,preprint,noshowpacs,tightenlines,eqsecnum,superscriptaddress,floatfix]{revtex4-1}
\usepackage{amsmath}
\usepackage{amssymb}
\usepackage{latexsym}
\usepackage{amsthm}
\usepackage{graphics}
\usepackage{enumerate}
\usepackage{bbm}
\usepackage{tensor}
\usepackage{amsmath}
\usepackage{amssymb}
\usepackage{tensor}
\usepackage{hyperref}
\newcommand{\be}{\begin{equation}}
\newcommand{\ee}{\end{equation}}
\newcommand{\ben}{\begin{equation*}}
\newcommand{\een}{\end{equation*}}
\newcommand{\bea}{\begin{eqnarray}}
\newcommand{\eea}{\end{eqnarray}}
\newcommand{\bean}{\begin{eqnarray*}}
\newcommand{\eean}{\end{eqnarray*}}

\newcommand{\bsub}{\begin{subequations}}
\newcommand{\esub}{\end{subequations}}

\newcommand{\disfrac}[1][2]{\displaystyle\frac}

\newcommand{\ima}{\mathbbmtt{i}}
\newcommand{\non}{\nonumber}
\newcommand{\bbar}{\overline}

\newtheorem*{theorem}{Theorem}
\def\bal#1\eal{\begin{align}#1\end{align}}
\begin{document}
\title{Decoupling of the re-parametrization degree of freedom and a generalized probability in quantum cosmology}
\author{N. Dimakis}
\email{nsdimakis@gmail.com}
\affiliation{Instituto de Ciencias Fisicas y Matematicas,\\
Universidad Austral de Chile, Valdivia, Chile}
\author{Petros A. Terzis}
\email{pterzis@phys.uoa.gr}
\affiliation{Nuclear and Particle Physics Section, Physics
Department,\\
University of Athens, GR 15771 Athens, Greece}
\author{Adamantia Zampeli}
\email{azampeli@phys.uoa.gr}
\affiliation{Nuclear and Particle Physics Section, Physics
Department,\\
University of Athens, GR 15771 Athens, Greece}
\author{T. Christodoulakis}
\email{tchris@phys.uoa.gr}
\affiliation{Nuclear and Particle Physics Section, Physics
Department,\\
University of Athens, GR 15771 Athens, Greece}

\begin{abstract}
The high degree of symmetry renders the dynamics of cosmological as well as some black hole spacetimes describable by a system of finite degrees of freedom. These systems are generally known as minisuperspace models. One of their important key features is the invariance of the corresponding reduced actions under reparametrizations of the independent variable, a fact that can be seen as the remnant of the general covariance of the full theory. In the case of a system of $n$ degrees of freedom, described by a Lagrangian quadratic in velocities, one can use the lapse by either gauge fixing it or letting it be defined by the constraint and subsequently substitute into the rest of the equations. In the first case, the system of the second order equations of motion is solvable for all $n$ accelerations and the constraint becomes a restriction among constants of integration. In the second case, the system can be solved for only $n-1$ accelerations and the ``gauge" freedom is transferred to the choice of one of the scalar degrees of freedom. In this paper, we take the second path and express all $n-1$ scalar degrees of freedom in terms of the remaining one, say $q$. By considering these $n-1$ degrees of freedom as arbitrary but given functions of $q$, we manage to extract a two dimensional pure gauge system consisting of the lapse $N$ and the arbitrary $q$: in a way, we decouple the reparametrization invariance from the rest of the equations of motion, which are thus describing the ``true" dynamics. The solution of the corresponding quantum two dimensional system is used for the definition of a generalized probability for every configuration $f^i (q)$, be it classical or not. The main result is that, interestingly enough, this probability attains its extrema on the classical solution of the initial $n$-dimensional system.
\end{abstract}
\maketitle
%---------------------------------------------------------------------------------------------------------------------------
%---------------------------------------------------------------------------------------------------------------------------

%---------------------------------------------------------------------------------------------------------------------------
%---------------------------------------------------------------------------------------------------------------------------
\section{Introduction}

One of the key elements of the theory of General Relativity is the diffeomorphism invariance of the action principle; one can map different line elements to each other by means of any (sufficiently smooth) invertible coordinate transformation without altering the geometry. This is in conjunction with the existence of (constraint) equations of motion involving only first derivatives of the metric and corresponds to the ``gauge'' freedom of choosing a coordinate system. In the Hamiltonian analysis of the full theory, where the configuration space becomes the space of all three-metrics (superspace),  generators of this freedom are the four constraints: the quadratic, together with the three linear in the momenta \cite{Dirac:1958sq}.

The presence of first-class constraints demands that symmetry considerations be taken into account. Symmetries play an important role in theoretical physics. In the case of cosmological models, one assumes a particular class of symmetries and proceeds with the reduction of the action from a field theory to a mechanical system. This procedure is called mini-superspace approximation and is an elegant way to truncate the theory to a system of finite degrees of freedom; in this context, symmetries of the configuration space have been considered. In particular, variational symmetries have been applied in the past (see e.g. \cite{Capozziello:1996ay}). Recently, the projective group was used in \cite{Tsamparlis:Geod,Tsamparlis:2011zz,MT2d} while in \cite{Christodoulakis:2013xha} the correspondence of the conditional symmetries \cite{Kuchar:1982eb} with the variational symmetries in a certain lapse parametrization was proven and consequently used for the study of minisuperspace models in \cite{Christodoulakis:2012eg,Christodoulakis:2013sya,Christodoulakis:2014wba,Dimakis:2013oza,Terzis:2014cra}.

If one considers a general line element of a spatially homogeneous Bianchi type
\be \label{lineel}
ds^2 = - N(t)^2 dt^2 + \gamma_{\mu\nu}(t)\sigma^\mu_i(x) \sigma^\nu_j(x) dx^i dx^j,
\ee
with the invariant basis one-form satisfying the relations $\sigma^\alpha_{i,j} -\sigma^\alpha_{j,i} = C^\alpha_{\beta \gamma} \sigma^\beta_j \sigma^\gamma_i$, then the Lagrangian function of the theory assumes the general form
\be \label{Lag}
L= \frac{1}{2 N} G_{\alpha\beta}(q)\dot{q}^\alpha \dot{q}^\beta - N V(q), \quad\quad \alpha,\,\beta = 0,\ldots, n-1,
\ee
where $N$ is the lapse function and $q^\alpha$ are the degrees of freedom that correspond to the components of the scale factor matrix $\gamma_{\mu\nu}$ and possibly any additional fields that one might add in the theory. One of course has to check whether the Euler - Lagrange equations corresponding to \eqref{Lag} generate the same dynamics as the field theory equations of motion evaluated for the above line element. When this is the case, \eqref{Lag} is considered valid. This Lagrangian has certain characteristic properties:
%-------------------------------------------------------------------------------------------------------------------------------

Let us write the corresponding action as
\begin{align}
I = \int dt N \left( \frac{1}{2} G_{\alpha \beta} \frac{d q^\alpha}{N dt} \frac{d q^\beta}{N dt}  -V\right) \label{action}
\end{align}
In this form, it can be easily seen that the action retains its form under reparametrizations of time and the corresponding changes of the dependent variables
\begin{align}
&t = f (\bar{t}), \\
&q^\alpha (t) = q^\alpha (f(\bar{t})) =: \bar{q}^\alpha (\bar{t}),\label{transformlaw} \\
&N(t) dt = N (f(\bar{t})) \frac{d f (\bar{t})}{d\bar{t}} d\bar{t} =: \bar{N} (\bar{t}) d\bar{t}.
\end{align}
The last transformation rule is evident from the line element \eqref{lineel} and designates $N$ as a density degree of freedom in contradistinction to $q^\alpha$'s which are scalars; when they refer to $\gamma_{\alpha\beta}$'s their transformation rule \eqref{transformlaw} is also evident from \eqref{lineel}. Alternatively, one could guess these transformations by just looking at the action and demanding that it remains form invariant. The above invariance is also present at the level of the equations of motion. Indeed, one can check that the Euler-Lagrange equations
\begin{align}
&E^0:= \frac{1}{2 N^2} G_{\alpha \beta} \dot{q}^\alpha \dot{q}^\beta + V =0, \label{constraint}\\
&E^\mu:= \ddot{q}^\mu + \Gamma^\mu_{\nu \lambda} \dot{q}^\nu
\dot{q}^\lambda -\frac{\dot{N}}{N} \dot{q}^\mu + N^2 G^{\mu \kappa} V_{,\kappa}\label{euler-lagrange}
\end{align}
do indeed respect this symmetry: by effecting the change $t= f(\bar{t})$, \eqref{constraint} is invariant, while \eqref{euler-lagrange} is multiplied by $(df/d\bar{t})^{-2}$ since the only troublesome term arising from the second derivative of time is counterbalanced by the third term containing $\frac{\dot{N}}{N}$.

A second important consequence that this invariance under arbitrary changes of $t = f(\bar{t})$ induces is that of the freedom to scale the lapse $N$ by an arbitrary function of the $q^\alpha$'s. In order to understand this property of the system it is useful to isolate the scalar degrees of freedom $q^\alpha$'s by solving \eqref{constraint} algebraically for $N^2$,
\begin{align}
N^2 = - \frac{G_{\alpha \beta} \dot{q}^\alpha \dot{q}^\beta}{2 V} =: -\frac{K}{2V} \label{lapse}
\end{align}
and then substituting it in \eqref{euler-lagrange} thereby obtaining
\begin{align}
\ddot{q}^\mu + \Gamma^\mu_{\nu \lambda} \dot{q}^\nu \dot{q}^\lambda -\frac{1}{2}\frac{\dot{K}}{K} \dot{q}^\mu + \frac{1}{2} \frac{V_{,\kappa}}{V} \dot{q}^\kappa \dot{q}^\mu - \frac{K}{2V} G^{\mu \kappa} V_{,\kappa}=0 \label{eqofmotion}
\end{align}
after replacing $\dot{V}= V_{,\kappa} \dot{q}^\kappa$. In this form, the equations are still covariant under arbitrary changes of time but they can only be solved for $n-1$ accelerations, as it can be seen by contracting with $2 G_{\mu \rho} \dot{q^\rho}$, thereby obtaining an identity $0=0$. At this stage, one can say that the freedom in defining the $t$-variable is translated into the freedom of arbitrarily choosing one of the $q^\alpha$'s. In this context, the constraint equation \eqref{constraint} simply specifies the lapse, once all the equations have been solved (after, of course, the arbitrary function has been specified).

Let us now take the Lagrangian \eqref{Lag} and consider a new independent variable $M= N e^{2 \omega}$, thereby obtaining the transformed Lagrangian
\begin{align}
\bbar{L} =\frac{1}{2M} \bbar{G}_{\alpha \beta} \dot{q}^\alpha \dot{q}^\beta - M\bbar{V}
\end{align}
where
\begin{equation}
\bbar{G}_{\alpha \beta} =e^{2 \omega} G_{\alpha \beta}, \qquad \bbar{V}=e^{-2 \omega} V. \label{transformations}
\end{equation}
By applying the same process on the equations of motion emanating from $\bbar{L}$ corresponding to $M$, $q^\alpha$'s we obtain the analogue of \eqref{eqofmotion}
\begin{align}
\ddot{q}^\mu + \bbar{\Gamma}^\mu_{\nu \lambda} \dot{q}^\nu \dot{q}^\lambda -\frac{1}{2}\frac{\dot{\bbar{K}}}{\bbar{K}} \dot{q}^\mu + \frac{1}{2} \frac{\bbar{V}_{,\kappa}}{\bbar{V}} \dot{q}^\kappa \dot{q}^\mu - \frac{\bbar{K}}{2\bbar{V}}\, \bbar{G}^{\mu \kappa}\, \bbar{V}_{,\kappa}=0. \label{eqofmotiontrasf}
\end{align}
This equation becomes identical to \eqref{eqofmotion} by using the transformations \eqref{transformations} and the well known relation
\begin{align}
\bbar{\Gamma}^\mu_{\nu \lambda} = \Gamma^\mu_{\mu \lambda} + \delta^\mu_\nu \omega_{,\lambda} + \delta^\mu_\lambda \omega_{,\nu} -G_{\nu \lambda} G^{\mu \rho} \omega_{,\rho}
\end{align}
This proves that one can scale $N$ arbitrarily and still solve the same equations containing accelerations. A particularly convenient, and useful, choice is to use this property in order to transform the potential $V$ into an effective constant potential $\bbar{V}$. Then, the dynamics is in a sense geometrized, since the Euler-Lagrange equations for the $q^\mu$'s become free geodesics in the scaled manifold's configuration metric (where of course the constraint selects the type of the geodesic vector).

Therefore, one is allowed to assume the transformation $N\mapsto \bbar{N} = N\, V$ first introduced in \cite{Misner:1972js} with the Lagrangian acquiring a constant potential
\be \label{Lagc}
\bbar{L}= \frac{1}{2\bbar{N}} \bbar{G}_{\alpha\beta}(q)\dot{q}^\alpha \dot{q}^\beta - \bbar{N}, \quad\quad \alpha,\,\beta = 0,\ldots, n-1,
\ee
with
\be \label{scmm}
\bbar{G}_{\alpha\beta}= V G_{\alpha\beta}
\ee
being the scaled mini-supermetric. Of course, as explained above, the dynamics before and after the scaling remains the same. For more information on the usefulness of this particular parametrization, especially at the quantum level, see \cite{Christodoulakis:2012eg,Christodoulakis:2013sya,Christodoulakis:2014wba,Zampeli:2015ojr}. From this point on we choose to work exclusively in the constant potential parameterization, thus we omit the bar symbolism.

As already mentioned, the time reparametrization invariance of the system makes it possible to solve the Euler - Lagrange equations of motion by leaving an unspecified degree of freedom, say $q^0$. If one wishes, it is possible to select this degree as representing the independent variable and describe the other dependent variables in terms of this, e.g. $q^i=f^i(q^0)$,  where $i=1,\ldots n-1$. This description is independent of any time reparametrization since all $q^\mu$'s transform, unlike the lapse function, as scalars \eqref{transformlaw}, thus leaving the relations $q^i=f^i(q^0)$ unaltered. The resulting solution of the Euler--Lagrange equations leads to a specific geometry for the base manifold which also satisfies the corresponding Einstein's equations and is designated by a particular set of $f^i$'s. One can then proceed to the Hamiltonian description and successively to the quantization of the system by enforcing the quantum analogues of the constraints as conditions on the wave function; likewise, any other observables that may be obtained due to the existence of symmetries (\cite{Christodoulakis:2012eg,Christodoulakis:2013sya,Jalalzadeh:2011yp,Vakili:2011nc,Vakili:2011uz,Capozziello:2012hm}). In the best case, a wave function can be completely determined by solving any existing eigen-equations together with the Wheeler-DeWitt constraint. However, this constitutes a quantization scheme that engulfs only the specific geometry under consideration.

\begin{figure}
\includegraphics{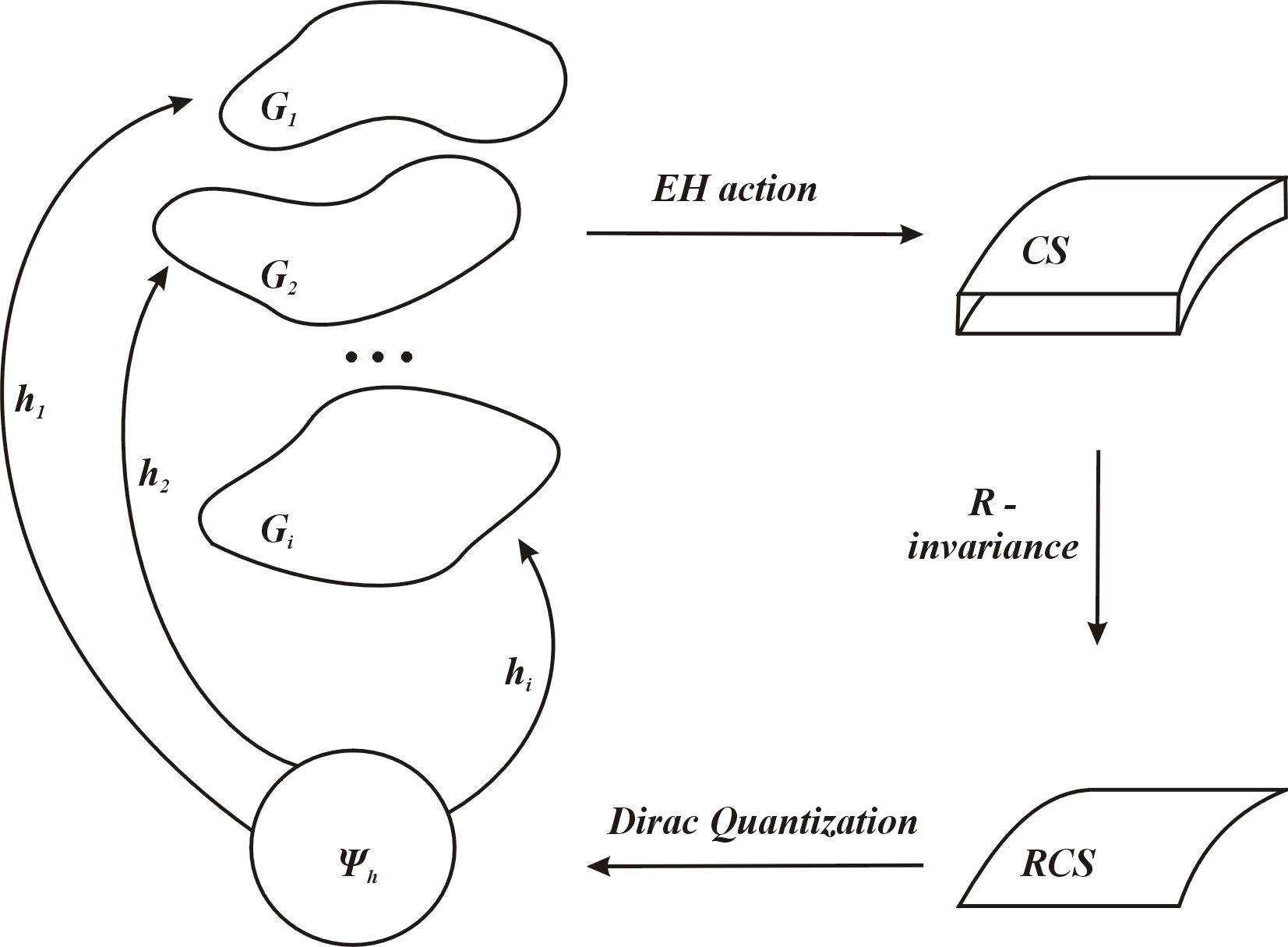}
\caption{The class of geometries $G_i$, define the configuration space (CS) through the minisuperspace Einstein--Hilbert action; the decoupling of the reparametrization invariance (R--invariance) leads to the reduced $(1+1)$ configuration space (RCS). The wave functional $\Psi_h$ which describes the class of geometries $G_i$ is obtained from the Dirac quantization of this system which finally gives a ``weight" $h_i$ to every geometry of the space time through the generalized probability \eqref{prob}.
}
\label{Procedure}
\end{figure}

In this work, we try to expand the quantization procedure given previously in the literature, in order to encompass all possible line elements inside a specific geometrical class. We use the information contained in the time reparametrization invariance - associated with the fact that not all degrees of freedom are independent - at the Lagrangian level. We thus perform a reduction from the initial system to another consisting of $1+1$ degrees of freedom; its dynamics is governed by two equations, one of which is of first order, i.e. a constraint, while the second order is sattisfied by virtue of the constraint. In this sense, we get a pure gauge system in which the previous degrees associated with scale factors of the base manifold have been reduced to arbitrary functions which can take any value. Consequently, the quantization of this latter system gives weight to all geometries (in a certain class) and surprisingly enough, the extrema of a generalized probability we define are located on the classical geometry corresponding to the initial system prior to the reduction. The whole procedure we follow can be visualized by FIG. \ref{Procedure}.

The paper is organized as follows: In section 2, we reduce the initial system by singling out one degree of freedom and in section 3 we proceed with the Hamiltonian formulation for the reduced system. In section 4, we perform its canonical quantization and we adopt a measure on the configuration space. This is necessary for the definition of a generalized probability function. In section 5, we study the conditions for the extrema of the generalized probability to exist, while section 6 is devoted to their characterization as minima or maxima. We conclude with two examples and some discussion of our results.
%-------------------------------------------------------------------------------------------------------------------------------%-------------------------------------------------------------------------------------------------------------------------------
\section{Decoupling of the initial system and its reduction}

We consider a system described by the Lagrangian \eqref{Lagc},
\be \label{lagpot}
L= \frac{1}{2\, N}\, G_{\alpha\beta}\dot{q}^\alpha \dot{q}^\beta - N, \quad \alpha,\beta= 0,\ldots n-1.
\ee
which corresponds to a mini-superspace analysis that was initiated by a line element of the form \eqref{lineel} in the constant potential parametrization.
The associated equations of motion are
\begin{subequations}\label{eqlag}
\begin{align} \label{singN}
2\, N^2 + G_{\alpha\beta}\, \dot{q}^\alpha\, \dot{q}^\beta & = 0\\ \label{doubledq}
\ddot{q}^\mu+ \Gamma^\mu_{\alpha\beta}\, \dot{q}^\alpha\, \dot{q}^\beta- \frac{\dot{N}}{N}\, \dot{q}^\mu &= 0.
\end{align}
\end{subequations}
Solving the constraint \eqref{singN} for $N^2$ and replacing in \eqref{doubledq}, we obtain
\begin{equation}
\ddot{q}^\mu + \Gamma^\mu_{\nu \lambda} \dot{q}^\nu \dot{q}^\lambda -\frac{1}{2} \frac{\dot{K}}{K} \dot{q}^\mu = 0  \label{constante-l}
\end{equation}
where $K$ is given by
\begin{align}
K=G_{\alpha \beta} \dot{q}^\alpha \dot{q}^\beta.
\end{align}
At this stage, the system is described by the equations \eqref{singN} and \eqref{constante-l}. We wish to disentangle the true dynamics from the gauge freedom $t = f (\bar{t})$. To this end, we choose one of the dependent scalar variables, say $q^0= q(t)$, and select the function $f$ as the inverse of $q$, $f:=q^{-1}$. Then, all of the $q^\alpha$'s can be expressed as functions of the selected $q$ and we are led to the definitions
\begin{subequations}\label{fis}
\begin{align}
& q^0=: q(t) = q(f (\bar{t})) = q (q^{-1} (\bar{t})) =\bar{t}, \label{fis1} \\
& q^i := q^i (t) = q^{i} (q^{-1}(\bar{t})) = q^i \circ q^{-1} (\bar{t}) =f^i(q), \quad i = 1,\ldots,n-1 . \label{fis2}
\end{align}
\end{subequations}
This means that we are effectively selecting $q(t)$ as time. As a result, the expression for $K$, using the obvious reletions $\dot{q
}^i (t) =\acute{f}^i (q(t)) \dot{q}(t)$ where $' = \frac{d}{dq}$, reads
\begin{align} \label{definitionh}
K =\left[ G_{00}+2 G_{0i} \acute{f}^i+G_{ij} \acute{f}^i\acute{f}^j \right] \dot{q}^2 =: h[q, f^i (q), \acute{f}^i (q)] \dot{q}^2.
\end{align}
Equation \eqref{definitionh} defines the functional $h[q, f^i (q) \acute{f}^i (q)]$, which, in the interest of concisness, we hereafter abbreviate as $h[q]$.

In order to simplify the forthcoming expressions, we adopt, without loss of generality, a Gauss-normal coordinate system in the configuration space, i.e. we assume coordinates in which the $(q, q^i), i=1,...,n-1$ are such that the metric elements are $G_{qq}=\varepsilon=\pm 1, G_{qi}=0$. Note that this singling out of a particular variable is not a feature of the selected coordinate system, but a general feature of the constrained systems as it has been discussed in detail in the introduction. Now we can express equation \eqref{singN} and decompose the equations \eqref{constante-l} in $\mu=0$ and $\mu=i$ components, with $\mu=0$ corresponding to the $q$ variable.
After the replacement of the non-vanishing Christoffel symbols, where $\Gamma^q_{jk} =-\frac{1}{2 G_{qq}} G_{jk,q}, \,\Gamma^i_{qk} = \frac{1}{2} G^{ij} G_{jk,q}$ and with $\Gamma^i_{jk}$ aquiring the regular form, as well as of the obvious relations $q'=1, \ q'' =0$, the quantity $K$ and the constraint assume the form
\begin{align}
K =\left[ \varepsilon+G_{ij} (q, f^i) \acute{f}^i (q)\acute{f}^j (q) \right]
\end{align}
\begin{equation}
N^2 = -\frac{h[q]}{2} \label{conh},
\end{equation}
while the $\mu=0$ component of \eqref{doubledq} takes the form
\begin{align}
\frac{h'}{h} = - \frac{1}{\varepsilon} G_{ij,q} \acute{f}^i \acute{f}^j \label{0-component}
\end{align}
and the $\mu=i$ component becomes
\begin{align}
\acute{\acute{{f}}}^i + G^{ik} G_{jk,q}\acute{f}^j + \Gamma^i_{kl} \acute{f}^k \acute{f}^l -\frac{1}{2} \frac{h'}{h} \acute{f}^i =0 . \label{extrema2}
\end{align}
Replacing \eqref{0-component} in \eqref{extrema2}, we get
\begin{align}
\acute{\acute{{f}}}^i +G^{ik} G_{jk,q}\acute{f}^j + \Gamma^i_{kl} \acute{f}^k \acute{f}^l +\frac{1}{2\varepsilon} G_{jk,q} \acute{f}^j \acute{f}^k \acute{f}^i =0, \label{accelerations}
\end{align}
a set of equations that are explicitly solved for the accelerations and describe the dynamics of the $f^i$'s degrees of freedom. These equations constitute a regular system with no other invariance under further arbitrary changes of the $q$ variable. An important property is that \eqref{extrema2} already contains the information of \eqref{0-component}; one can see this by contracting \eqref{extrema2} with $2 G_{ir} \acute{f}^r$ and using the derivative of the definition of $h$ to replace the term $2 G_{ir} \acute{f}^r \acute{\acute{{f}}}^i$. Thus the initial system of equations \eqref{eqlag}, in which the dependent variables $N, q^\alpha$ were coupled, has now been transformed into equations \eqref{conh} and \eqref{accelerations}. In other words, the system is decoupled in the variables  $N, q$ and $ f^i (q)$.

We now pose the question: what is the dynamics of the $N,q$ degrees of freedom when $f^i(q)$ are specified albeit arbitrary functions, not necessarily obeying \eqref{accelerations}, so that we can describe the entire family of geometries of a given base manifold's line element. To unravel the answer, it is sensible to turn to the Lagrangian \eqref{lagpot} and replace the $q^i$'s from \eqref{fis2} and their time derivatives by $\dot{q}^i (t) =\acute{f}^i (q(t)) \dot{q}(t)$. This action defines a new Lagrangian which - among other properties - could, upon quantization, be capable of associating a probability to each and every geometry in a given class defined by $f^i (q)$'s. The reduced Lagrangian takes the form
\be \label{lagred}
L_{red} = \frac{1}{2\, N(t)}\, h[q(t)]\, \dot{q}(t)^2- N(t),
\ee
and describes a constrained system with two degrees of freedom, $N$ and $q$.

The reduced Lagrangian \eqref{lagred} is radically different from the  original Lagrangian \eqref{Lag}, since the former contains the \emph{arbitrary} functions $f^i(q(t))$ (not to be considered as degrees of freedom), whilst the latter is a ``usual'' Lagrangian with the $q^\alpha(t)$ playing the role of the ordinary degrees of freedom. Thus the equations of motion that result from the reduced Lagrangian are ordinary differential equations with \emph{arbitrary} functions in them; therefore these equations describe a \emph{class} of geometries. This fact is made more clear in the treament of the example of the static and spherically symmetric spacetime in section \ref{secexamples}.

The equations of motion for the above Lagrangian are:
\begin{subequations} \label{eulred}
\begin{align} \label{eulredn}
E^0_{red} &:= 2\, N^2 + h[q]\, \dot{q}^2 = 0 \\ \label{eulredq}
E_{red} &:= \ddot{q} + \frac{h'}{2\, h} \, \dot{q}^2 - \frac{\dot{N}}{N}\, \dot{q}=0,
\end{align}
\end{subequations}
It can be easily verified that equation \eqref{eulredq} is a consequence of \eqref{eulredn}: by solving the former for $N$ and substituting in the later we get that its left hand side becomes identically zero. This is not just a property of the system under consideration. Every constrained system of $1+1$ degrees of freedom has its dynamics determined by only the first order constraint equation. Therefore, one can either regard the constraint as a first order differential equation for $q$ (when the lapse is to be prescribed), or as an algebraic equation for the lapse. The solution of the constraint leaves the other degree of freedom arbitrary. Thus, the equation of motion for $q(t)$ is satisfied by virtue of the constraint, without $q(t)$ itself being explicitly determined and, more importantly, without any conditions being imposed on the $f^i(q)$'s. Any properties of the reduced system defined by \eqref{lagred} hold irrespectively to the particular form of the $n-1$ functions $f^i(q)$. Additionally, we keep in mind that some or all of the $f^i(q)$'s are components of the base manifold metric \eqref{lineel}. Thus, the properties of the reduced system hold for any particular geometry contained in the general form of \eqref{lineel} (for fixed $\sigma^\alpha_i(x)$); a key reason for considering this reduced Lagrangian.

%-------------------------------------------------------------------------------------------------------------------------------%-------------------------------------------------------------------------------------------------------------------------------
\section{Phase space description}

The conjugate momenta for the reduced system are:
\begin{subequations}
\begin{align}
p_N & := \frac{\partial L_{red}}{\partial \dot{N}}=0, \\
p &:= \frac{\partial L_{red}}{\partial \dot{q}}= \frac{h}{N}\, \dot{q}.
\end{align}
\end{subequations}
By following Dirac's prescription for singular systems (\cite{Dirac:1958sq,Dirac:1950pj,Dirac:113811,Sundermeyer:1982gv}), we recognize $p_N\approx 0$ as the primary constraint, thus the Hamiltonian is written as
\begin{align} \nonumber
H_T & = \dot{q}\, p - L + u^N\, p_N \\ \nonumber
  & = N\left(\frac{p^2}{2\, h}+1\right)+ u^N\, p_N \\
  & = N\, \mathcal{H} +  u^N\, p_N,
\end{align}
where we have set
\be \label{seccon}
\mathcal{H}= \frac{p^2}{2\, h}+1 .
\ee

The primary constraint must be preserved in time, i.e.
\be
\dot{p}_N = \{p_N, H_T\} \approx -\mathcal{H} \approx 0.
\ee
Consequently, we are led to the secondary constraint $\mathcal{H}$ which produces no tertiary constraints. Obviously, the two constraints are first class, $\{p_N,\mathcal{H}\}\approx 0$, meaning that their coefficients in the Hamiltonian are not fixed by the theory, since the Dirac-Bergmann algorithm for finding all the constraints is being terminated. From the secondary constraint, we can easily deduce that
\be \label{intmoeasyway}
\mathcal{H} \approx 0 \Rightarrow \frac{p^2}{-h}\approx 2 \Rightarrow \frac{p}{\sqrt{-h}} = \pm \sqrt{2},
\ee
which means that $Q=\frac{p}{\sqrt{-h}}$ is a linear in the momenta integral of motion. Note here that the same result can be obtained formally: if we set $Q=A(q)p+B(q)$, where $A(q)$, $B(q)$ are arbitrary functions, and demand this quantity to be a conditional symmetry \cite{Kuchar:1982eb} (integral of motion by virtue of the constraints, i.e. $\approx 0 \Rightarrow \; = F(q)\mathcal{H}$, with $F(q)$ an arbitrary function), we get
\begin{align*}
\dot{Q} \approx 0\Rightarrow \{Q,H_T\} &= N\, F(q)\, \mathcal{H} \Rightarrow \\
\frac{N}{h}\left(2\, A'+ A\, \frac{h'}{h}\right)\, p^2 + \frac{2\, N\, B'}{h}\, p & = \frac{F\, N}{h}\, p^2+F\, N.
\end{align*}
and by gathering the coefficients of different powers of the momentum $p$ we deduce
\begin{align} \nonumber
B(q) &= \text{constant}\\ \nonumber
F(q) & = 0 \\
2\, A' + A\, \frac{h'}{h} &=0,
\end{align}
with the last equation having as a solution $A=\frac{c_1}{\sqrt{-h}}$. The general form of the linear integral of motion is thus $Q=\frac{c_1\, p}{\sqrt{-h}}+ c_2$. As it can be seen by the form of \eqref{intmoeasyway}, the presence of $c_1$ corresponds to the multiplication of the Hamiltonian with an overall constant, while that of $c_2$ presents the trivial addition of any constant to an integral of motion; thus we restrict ourselves to the particular form
\be \label{intmo1d}
Q=\frac{p}{\sqrt{-h}}.
\ee
It is important to note once more that the functions $f^i(q)$ have not suffered the imposition of any conditions. The quantity \eqref{intmo1d} is constant whenever the Euler-Lagrange equations \eqref{eulred} are satisfied and this happens for arbitrary $f^i(q)$. Consequently, $Q$ is an integral of motion of \eqref{lagred} irrespective to the particular geometry that the non reduced system \eqref{Lag} might have as a solution.
%-------------------------------------------------------------------------------------------------------------------------------%-------------------------------------------------------------------------------------------------------------------------------
\section{Quantization of the reduced system}

We proceed with the quantization of the system by adopting the usual canonical commutation relations. The momenta operators are taken to be
\begin{subequations}
\begin{align}
\widehat{p}_N &= - \ima \frac{\partial}{\partial N} \\
\widehat{p} &= - \ima \frac{\partial}{\partial q}
\end{align}
\end{subequations}
and the Poisson brackets are mapped to the commutator as
\be
\{\;,\;\}\rightarrow -\ima[\;,\;],
\ee
where for brevity we have adopted the units $\hbar=1$.

Imposition of the primary constraint on the wave function, $\widehat{p}_N\Psi=0$, leads to $\Psi=\Psi(q)$. Thus, $\Psi$ becomes a function of a single variable. For the operator counterpart of \eqref{intmo1d}, we choose the most general form of an operator Hermitian under an arbitrary measure $\mu(q)$ \cite{Christodoulakis:1990zy},
\be \label{qhat1d}
\widehat{Q} = - \frac{\ima}{2\, \mu}\left(\frac{\mu}{\sqrt{-h}}\,\frac{d}{d q}+\frac{d}{d q}\left(\frac{\mu}{\sqrt{-h}}\right)\right).
\ee
Since $Q$ corresponds to a classical constant of motion, we infer that at the quantum regime the eigenvalue equation
\be \label{qcon1d}
\widehat{Q}\Psi = \kappa\Psi
\ee
where $\kappa= \pm \sqrt{2}$ must be enforced.

For the quadratic constraint, we use the operator
\be
\widehat{\mathcal{H}} = -\frac{1}{2\, \mu}\, \frac{d}{d q} \left(\frac{\mu}{h}\,\frac{d}{d q}\right)+1,
\ee
whose kinetic part becomes the one dimensional Laplacian if $\mu=\sqrt{-h}$ (natural measure) and which, in general, has the property that when $\mu$ transforms like a scalar density, $\widehat{\mathcal{H}}$ transforms as a scalar \cite{Christodoulakis:1990zy}. For the time being we consider the measure arbitrary. Given the previous definitions, the Wheeler-DeWitt equation reads
\be \label{WDW1d}
\widehat{\mathcal{H}}\Psi =0 \Rightarrow \frac{\Psi''}{2\, h}+ \frac{h\, \mu' -\mu\, h'}{2\, h^2\, \mu}\, \Psi' - \Psi =0.
\ee

Equations \eqref{qcon1d} and \eqref{WDW1d} are to be satisfied by the wave function we seek. The eigenvalue equation \eqref{qcon1d} can, for an arbitrary measure, be easily integrated yielding
\be \label{sol1d}
\Psi_h [q]= \frac{(-h[q])^{\frac{1}{4}}}{\sqrt{\mu(q)}}\, e^{\ima\, \kappa \, \int\!\! \sqrt{-h[q]}\,d q}.
\ee
Of course, it is left for us to check what conditions \eqref{WDW1d} imposes on this solution. To this end, we insert the result \eqref{sol1d} into the Wheeler-DeWitt equation and, for the sake of simplicity, we set $\mu[q] = \phi[q]^2\, \sqrt{-h[q]}$; the outcome, under the classical restriction $\kappa = \pm \sqrt{2}$, is the following expression for $\phi [q]$
\be
h'\, \phi' - 2\, h\, \phi'' = 0 \Rightarrow \phi[q] = c_1 + c_2 \int\!\! \sqrt{-h[q]}\, d q ,
\ee
which sets a condition on the measure,
\be \label{genmeas1d}
\mu[q]= \sqrt{-h[q]}\left(c_1+c_2 \int\sqrt{-h[q]}\, d q\right)^2.
\ee
As one can see, the natural measure is contained in this solution ($c_1=1$, $c_2=0$). Had we originally adopted the physical measure, the Wheeler-DeWitt equation would have become an identity after the imposition of the eigenvalue equation. Additionally, one can easily check that, under the general measure \eqref{genmeas1d}, the relation
\be\label{Qevolution}
[\widehat{Q},\widehat{\mathcal{H}}]=0
\ee
holds in general, not only on the wave function $\Psi$ \footnote{The relation \eqref{Qevolution} is also satisfied for solutions of \eqref{WDW1d} that are not necessarily on-mass shell, i.e. for which $\kappa \neq \pm \sqrt{2}$ and the measure is not given by the form \eqref{genmeas1d}.}. However, the particular form of $\mu(q)$ is not going to concern us any more, because the corresponding probability that one can define as
\be
\mu[q]\, \Psi[q]^* \, \Psi[q] dq = \sqrt{-h[q]} dq
\ee
is independent of it.

Due to the form of \eqref{sol1d} and since we know from the theory of quantum mechanics that the different states are distinguised by generically different (not up to a constant) phases, we can consider a generalized probability amplitude for a state of the system $h$ to be given by
\be \label{prob}
P_{h} = \int\!\! \sqrt{-h[q]}\, d q.
\ee
The above generalized probability is a step towards the probabilistic description of the quantum system (spacetime geometries) described by the wave functional $\Psi_h[q]$. The basic difference with the standard probability is that \eqref{prob} is not normalized to take values in $[0,1]$; in fact, the presence of the unspecified $f^i $'s prohibits any attempt to check normalizability (for unormalized or relative probability see e.g. p.73 of \cite{dirac1958principles}). However, this disadvantage is counterbalanced by the benefits of its introduction:

\begin{enumerate}
\item The use of the generalized probability leads naturally to the connection of the quantum and classical treatment of a spacetime geometry. To the best of our knowledge this is the first time that someone can give a physical meaning to the wave functional of the space time $\Psi_h[q]$. Until now every information gained from the solution of the Wheller--DeWitt equation was restricted in the context of the semiclassical analysis, e.g. \cite{Hartle:1983ai,Christodoulakis:2013sya}. Now the choice of a particular configuration $q^i(t)$ in the base manifold connects directly with $\Psi_h[q]$ through the functional $h[q]$.

\item As a consequence of the above, \emph{every} configuration $q^i(t)$ acquires a weight through the probability \eqref{prob}. The dominant argument for announcing $P_h$ as a probability is that it attains its extreme value when $q^i(t)$ correspond to the classical solution, as it will be shown in the next section.
\end{enumerate}
%----------------------------------------------------------------------------------------------------------------
%----------------------------------------------------------------------------------------------------------------

\section{Candidates for extrema of the generalized probability $P_h$}\label{candidates}

In order to check for the extrema of the functional \eqref{prob}, we are obliged to consider the functions  $f^i (q)$ implicit in the functional $h[q]$ as eligible for variation,
\be
\delta P_{h} = \delta \int\!\!\sqrt{-h[q]}\, d q = 0,
\ee
with
\be
\sqrt{-h[q,f,\acute{f}]}= \left(-G_{ij}(q)\,\acute{f}^i(q)\, \acute{f}^j(q)- \varepsilon \right)^{\frac{1}{2}}.
\ee
For simplicity, from now on we omit the minus sign inside the root (since $P_{-h}$ exhibits the same extrema as $P_{h}$). The demand that $\delta P_{h}=0$ leads to the set of equations
\begin{align}
\acute{\acute{f}}^m +  G^{mk} G_{rk,q} \acute{f}^r +\Gamma^m_{rn} \acute{f}^r \acute{f}^n  -\frac{\acute{h}}{2 h} \acute{f}^m =0 \label{extrema1}
\end{align}
This proves that the extrema of the generalized probability lie on the classical solution of the initial Lagrangian, since \eqref{extrema1} is identical to \eqref{extrema2} and therefore equivalent to \eqref{accelerations} which describes the classical dynamics. Hence, we can state the following:
\begin{theorem}
Given a cosmological mini-superspace model with a valid Lagrangian of the form
\be
L = \frac{1}{2\, N}\, G_{\alpha\beta}\, \dot{q}^\alpha\, \dot{q}^\beta- N\, V, \quad\quad \alpha, \beta=0,\ldots n-1,
\ee
the possible existing extrema for the generalized probability
\be
P_{h} = \int\!\! \sqrt{-h}\, d q ,
\ee
with $h:=\left(V\, G_{\alpha\beta}\dot{q}^\alpha\dot{q}^\beta\right)\big|_{q=t}$ and $t$ the dynamical independent variable of the system, rest at the classical solutions of Einstein's equations.
\end{theorem}
%--------------------------------------------------------------------------------------------
This is the main result of this work. In order to make the application of the above theorem more concrete, we analyze the examples of a static, spherically symmetric metric and a spatially homogeneous one in section \ref{secexamples}.

Quantization of the reduced system, whose $q^i=f^i(q)$ do not acquire any particular values, encompasses all possible geometries spanning the base manifold. The generalized probability amplitude obtained herein reveals as its possible extrema the classical solutions of the initial system. By this method the reduced system gives weight to all the geometries contained in a certain class, but distinguishes the classical configurations. However, it must be noted that different functions $h(q,f,\acute{f})$ may have their extrema on the same base manifold geometry, since the $f^i(q)$ can be re-parameterized with respect to each other or because a different gauge fixing condition might be taken into consideration. In other words, what we call here as state $h[q]$ corresponds to a specific parametrization of the initial configuration and it is thus possible for different states to exhibit their extrema on the same classical geometry.

Finally we would like to point out that the quantity $\sqrt{-V G_{\alpha\beta}q'^\alpha q'^\beta}$ with $\alpha,\beta$ taking values from $0$ to $n-1$ and $G_{\alpha\beta}$ being the unscaled mini-supermetric can be seen as a reduced form of the Baierlein-Sharp-Wheeler (BSW) action \cite{Baierlein:1962zz}:
\be \label{BSWact}
S_{BSW} = \int\!\! \sqrt{g}\,\sqrt{R}\,\sqrt{k^{ij}k_{ij}-\mathrm{tr}k^2}\,d^4 x
\ee
where $g=\det g_{ij}$ is the determinant of the spatial metric and $k_{ij}=-2\,N\,K_{ij}=\frac{\partial g_{ij}}{\partial t} - N_{i;j} - N_{j;i}$, with $K_{ij}$ being the extrinsic curvature. It is known that \eqref{BSWact} can be used to produce the same dynamics as the Einstein-Hilbert action (for more details see \cite{Barbour:2000qg,2002Murchadha}).

The fact that $\delta \int{\sqrt{-h}} dq=0$ leads to the equations of motion for the system described by \eqref{lagpot} is in line with the property of the BSW action, even though the BSW action enjoys the reparametrization invariance, while the generalized probability does not. It is interesting to note that, in the semiclassical analysis, the wave function is taken to be $\Psi \propto e^{\ima S}$, where $S$ is the action that describes the classical system integrating over all possible metrics \cite{Hartle:1983ai}. In the quantum description of the reduced system that we present here, the wave function for a given state becomes $\Psi \propto e^{\ima S_{BSW}}$ in a natural way, i.e. as the combined solution of the system consisting of the eigenfunction equation \eqref{qcon1d} and the quadratic constraint \eqref{WDW1d} \cite{Carlini:1995gn}.

%----------------------------------------------------------------------------------------------------------------
%----------------------------------------------------------------------------------------------------------------
%-----------------------------------------------------------------------------------------------------------------

%----------------------------------------------------------------------------------------------------------------
\section{Characterization of the candidates for extrema}

The characterization of the candidates for extrema as either maxima or minima for a given functional is a rather tedious and complicated task that depends on many factors (among them even the range of integration). We refer the interested reader to various textbooks on the subject (for example \cite{gelfand1964calculus}, \cite{van2003calculus}). We briefly state some basic facts of the theory; one has to proceed with the second variation of the functional and check whether, on the extremizing configurations, has positive (minimum) or negative (maximum) value. In our case, this becomes
\be \label{secv}
\delta^2 \sqrt{-h}= \frac{\partial^2 \sqrt{-h}}{\partial \acute{f}^i \partial \acute{f}^j} \, \delta \acute{f}^i \, \delta \acute{f}^j+2\, \frac{\partial^2 \sqrt{-h}}{\partial \acute{f}^i \partial f^j} \, \delta \acute{f}^i \, \delta f^j + \frac{\partial^2 \sqrt{-h}}{\partial f^i \partial f^j} \, \delta f^i \, \delta f^j .
\ee
Then, by writing the second term as
\be
2\, \frac{\partial^2 \sqrt{-h}}{\partial \acute{f}^i \partial f^j}\, \delta \acute{f}^i \, \delta f^j = \frac{d}{dq} \left(\frac{\partial^2 \sqrt{-h}}{\partial \acute{f}^i \partial f^j}\, \delta f^i \, \delta f^j\right) - \frac{d}{dq} \left(\frac{\partial^2 \sqrt{-h}}{\partial \acute{f}^i \partial f^j}\right)\, \delta f^i \, \delta f^j ,
\ee
and using the requirement that the variation $\delta f^i$ should be zero on the boundary (together with the assumption of a well behaved term $\frac{\partial^2 \sqrt{-h}}{\partial \acute{f}^i \partial f^j}$), \eqref{secv} becomes
\be \label{deltasqh}
\delta^2 P_{h}= \int_a^b\!\! \left[\frac{\partial^2 \sqrt{-h}}{\partial \acute{f}^i \partial \acute{f}^j}\, \delta \acute{f}^i \, \delta \acute{f}^j + \left(\frac{\partial^2 \sqrt{-h}}{\partial f^i \partial f^j}- \frac{d}{dq}\left(\frac{\partial^2 \sqrt{-h}}{\partial \acute{f}^i \partial f^j}\right)\right)\, \delta f^i \, \delta f^j\right]\, dq.
\ee
In the test for the characterization of the candidates for extrema, the following two symmetric matrices are important
\begin{subequations} \label{matrWA}
\begin{align}
W_{ij} & = \frac{\partial^2 \sqrt{-h}}{\partial \acute{f}^i \partial \acute{f}^j} \\
A_{ij} & = \frac{\partial^2 \sqrt{-h}}{\partial f^i \partial f^j}- \frac{d}{dq}\left(\frac{\partial^2 \sqrt{-h}}{\partial \acute{f}^{(i} \partial f^{j)}}\right).
\end{align}
\end{subequations}
According to the theory, one has first to secure a matrix $U$ that is a valid solution of the following Riccati equation between matrices
\be \label{matric}
A + \acute{U} = U\, W^{-1}\, U
\ee
in the region of integration $[a,b]$, i.e. it is well behaved in this region. Given the existence of such a $U$, an adequate total derivative term can be added to the right-hand side of \eqref{deltasqh} (on account of the boundary condition), thus making the sign of $\delta^2 P_{h}$ dependent on whether the matrix $W$ is positive or negative definite. In the first case, the solution of $\delta P_{h}=0$ is a minimum of the functional, while in the latter it is a maximum. There is also the possibility that $W$ is neither positive nor negative definite, in this instance the test is considered inconclusive.

In our case the coefficients of the matrix $W$ read
\begin{align} \nonumber
W_{ij} & = \frac{\partial^2 \sqrt{-h}}{\partial \acute{f}^i \partial \acute{f}^j} = \frac{\partial}{\partial \acute{f}^i} \left(-\frac{1}{2\, \sqrt{-h}}\frac{\partial}{\partial \acute{f}^j}\left( \varepsilon+ G_{kl}\, \acute{f}^k\, \acute{f}^l\right)\right) \\ \nonumber
& = \frac{\partial}{\partial \acute{f}^i} \left(-\frac{1}{\sqrt{-h}}\, G_{kj}\, \acute{f}^k\right) \\ \nonumber
& =- \frac{G_{ik}\,G_{jl}\acute{f}^k\, \acute{f}^l}{(-h)^{\frac{3}{2}}}+ \frac{G_{ij}}{\sqrt{-h}} \\ \label{Wij}
& = - \frac{1}{\sqrt{-h}} \left(\frac{1}{-h}G_{ik}\,G_{jl}\acute{f}^k\, \acute{f}^l+G_{ij}\right),
\end{align}
where the $f^i$'s are to be substituted from the solution of \eqref{extrema1}.

%-------------------------------------------------------------------------------------------------------------------------------%-------------------------------------------------------------------------------------------------------------------------------
\section{The Schwarzschild and the Bianchi Type III states}\label{secexamples}
The adoption of a static, spherically symmetric line element of the form
\be \label{stat}
ds^2 = - a(r)^2 dt^2 + \left(\frac{N(r)}{2a(r)}\right)^2 dr^2 +  b(r)^2 (d\theta^2+ \sin^2\theta\,d\phi^2 )
\ee
where the radial coordinate $r$ is taken as the evolution parameter of the system, leads to the following mini-superspace Lagrangian in the case of vacuum
\be \label{ScwLag}
L =  -\frac{1}{2\, N}\left(16\, a\, b\, \dot{a}\, \dot{b} + 8\, a^2\, \dot{b}^2 \right)-N
\ee
where now $\dot{}=\frac{d}{dr}$. Line element \eqref{stat} is of the form \eqref{lineel} with $r$ being the dynamical parameter instead of $t$. By adopting this $r$-evolution, a valid mini-superpace can be given (a procedure first introduced in \cite{Cavaglia:1994yc} and \cite{Cavaglia:1995bb} and later used in \cite{Jalalzadeh:2011yp,Vakili:2011nc,Christodoulakis:2012eg,Christodoulakis:2013sya}). Note that in \eqref{stat} we already parametrized the ``lapse" function in the line element as $\frac{N(r)}{2a(r)}$, so that the ensuing potential would be constant. Equivalently, we could set the coefficient of $dr^2$ as $N(r)^2$ and then re-scale the lapse inside the Lagrangian $N\mapsto \frac{\bbar{N}}{2a}$.

One can easily check that the Euler - Lagrange equations of \eqref{ScwLag} lead to the well known Schwarzschild solution
\be \label{ScwSol}
ds^2 = -c^2\left(1-\frac{2M}{b(r)}\right)dt^2 + \left(1-\frac{2M}{b(r)}\right)^{-1} \dot{b}(r)^2 dr^2  +b(r)^2 \left(d\theta^2+ \sin^2\theta\,d\phi^2\right).
\ee
Note that this particular form of the line-element has been acquired without any gauge fixing, hence the existence of an arbitrary function $b(r)$.

In order to keep in touch with the classical solution \eqref{ScwSol}, we choose to construct the reduced system by choosing the scale factor $b(r)$ as our dynamical independent variable. Then, according to $h[q]$ and given that the mini-supermetric induced from \eqref{ScwLag} reads
\be
\label{sgbar}
\bbar{G}_{\alpha\beta} =
\begin{pmatrix}
 0 & -8\, a\, b \\ \\
 -8\, a\, b & -8 \, a^2
 \end{pmatrix},
\ee
we deduce that the $h$ function in this case is
\be
h_S[b] = - 8\, \left(2\, b\, a(b)\, a'(b) + a(b)^2 \right).
\ee
As expected, the classical reduced system does not choose any particular form for the $a(b)$. The Euler-Lagrange equation of
\be
L=\frac{1}{2N}h_S[b] \dot{b}^2 -N
\ee
for $b$ is satisfied whenever the corresponding equation for $N$ is satisfied.

The Euler--Lagrange equation for $N(r)$ reads
\bal\label{EL_N(r)}
b'(r)^2\left(b a(b) \right)'=\frac{1}{4}N(r)^2,
\eal
thus adopting the gauge $N(r)dr=2d\tau$, i.e $N(r)\mapsto 2$, we have the general solution of \eqref{EL_N(r)} in parametric form
\bal\label{sol_Par}
\tau		&=\rho f''(\rho)-f'(\rho) \non\\
b 		&=f''(\rho) \non\\
a^2		&=\frac{1}{f''(\rho)}\left( \rho^2 f''(\rho)-2\rho f'(\rho)+2f(\rho) \right)
\eal
where $f(\rho)$ is an arbitrary function. The above solution has no contact with the solution of the original Lagrangian \eqref{ScwLag}, since the original degree of freedom $a(r)$ has transformed into the arbitrary function $a(b)$. Eventually the arbitrariness of the function $a$ is transferred to the arbitrariness of the function $f$.

The generalized probability in our case is defined as
\be \label{genprob}
P_{h_S}= \int\!\! \sqrt{-h_S[b]}\, d b
\ee
and the value of $a(b)$ for which it may exhibit an extrema is given by the solution of
\be
\frac{d}{d b}\left(\frac{\partial \sqrt{-h}}{\partial a'}\right)- \frac{\partial  \sqrt{-h}}{\partial a} = 0 \Rightarrow (b\, a'' + 2\, a')\, a + b\, a^2 =0
\ee
The latter can be easily found to be
\be \label{orbitab}
a_S(b) = c\, \sqrt{1-\frac{2\, M}{b}},
\ee
which is identical to the classical solution for the scale factor, see line element \eqref{ScwSol}

In order to proceed with the characterization of $a_S$, we need to calculate the matrices $W$ and $A$ \eqref{matrWA}. Here, due to the fact that there is only one $f^i(q)$, that is $a(b)$, we obtain the functions
\begin{subequations}
\begin{align}
W(b)  = & \frac{\partial^2 \sqrt{-h}}{\partial a' \partial a'} = - \frac{2\, \sqrt{2}\, b^2\, a^2}{\left(a\, (a+2\, b\, a')\right)^{\frac{3}{2}}} \\ \nonumber
A(b)  = & \frac{\partial^2 \sqrt{-h}}{\partial a \partial a} - \frac{d}{d b} \left(\frac{\partial^2 \sqrt{-h}}{\partial a \partial a'}\right) \\
= & - \frac{2\,\sqrt{2}\, b\, a \left(b^2\, a^{\prime 3}-b^2\, a\, a'\,a'' + a^2\, (2\, a' + b\, a'')\right)}{\left(a\, (a+2\, b\, a')\right)^{\frac{5}{2}}} .
\end{align}
\end{subequations}
These, on the solution $a=a_S$, assume the following values:
\begin{subequations}
\begin{align} \label{wab}
W\big|_{a =a_S} (b) = & -\frac{2\, \sqrt{2}\, b^2\, (1-\frac{2\, M}{b})}{|c|}, \\
A\big|_{a =a_S} (b) = & -\frac{2\, \sqrt{2}\, M^2}{|c|\, b\, (b - 2\, M)}.
\end{align}
\end{subequations}
As we also mentioned in the previous section the Riccati equation
\begin{align} \nonumber
& A\big|_{a =a_S}(b) + U'(b) = \frac{1}{W\big|_{a =a_S} (b)}\, U(b)^2 \Rightarrow \\ \label{Ric}
& 2\,\sqrt{2}\, c \, (b-2\, M)\, U'(b) + c^2\, U(b) -8\, M^2 =0
\end{align}
must have a well behaved solution $U(b)$ over the region of integration of the functional $P_{h_S}$. Indeed one can integrate \eqref{Ric} to find
\be \label{solU}
U(b) = \frac{2\, \sqrt{2}\, M}{c}\, \tanh\left(\frac{1}{2}\left(\mathrm{ln}|b-2\, M|-\mathrm{ln} b-4\, \sqrt{2}\, c\, M\, c_1 \right)\right),
\ee
with $c_1$ being the constant of integration. We observe that \eqref{solU} is well behaved in $\mathbb{R}$, even at $b=0$ and $b=\pm 2\, M$ which at the classical level correspond to a curvature and a coordinate singularity respectively. Due to the fact that $\tanh(x)$ is bounded, \eqref{solU} is well behaved even when $b\rightarrow \pm\infty$. Thus, we can continue with the characterization of $a_S(b)$.

If we assume that we may, at the quantum level, correlate $b$ with the radial parameter of the classical regime, we can see from \eqref{wab} that the sign of $W\big|_{a =a_S} (b)$ does not remain constant in the region $(0,\infty)$. Outside the Schwarzschild horizon $b>2\, M$, the generalized probability assumes its maximum value when $a=a_S$. On the contrary, inside the horizon $b<2\, M$ the situation is exactly the opposite and $P_{h_s}$ takes its least value. Consequently, we may somewhat heuristically observe that the measurement outside the horizon is most likely to show that the space-time is characterized by the Schwarzschild metric, if we suppose that one can ``measure" geometry. On the other hand, inside the horizon there is no particular geometry that emerges as more probable. We can only say that the least probable is the one given by line element \eqref{ScwSol}.
%----------------------------------------------------------------------------------------------------------------

Similar results are obtained in the case of the LRS Bianchi Type III cosmological model with spacetime metric
\begin{align}
ds^2 = -\left( \frac{N(t)}{2\, a(t)}\right)^2 dt^2 + b(t)^2 dx^2 + e^{-2x} b(t)^2 dy^2 + a(t)^2 dz^2
\end{align}
i.e.
As a result, the generalized probability \eqref{genprob} results in a similar form of its extremum
\begin{equation}
a(b) = c_2 \sqrt{1-\frac{c_1}{b}}
\end{equation}
where $c_1$ is an essential constant. When $b>c_1$, we have the maximum and the time parameter is $t$, while for $b<c_1$ we obtain the minimum value and $z$ becomes the time parameter with the appropriate redefinitions of the constants. In this case, the space fails to be spatially homogeneous any more.

A rather interesting fact that can be pointed out is that the minisupermetrics of these two examples are also flat; therefore, there exists a transformation of the configuration variables that maps one to the other. These two examples profoundly demonstrate the case in which the function $h$ is the same for two spacetimes not belonging to the same geometrical class. However, if the two line elements are given, the difference in the solutions can be inferred by their invariant relations
\begin{align}
I_{Schw} &= q_2 - \frac{2^{2/3} \, 3^{5/3} q_1}{m^{2/3}} -6 \sqrt{3} q_1^{5/2}, \\
I_{III} &= q_2 -  (6 \sqrt{3} q_1^{5/2} -3 \times 6^{2/3} c_1^{2/3} q_1^{7/3})
\end{align}
where $q_1 $ is the Kretchmann scalar and $q_2 = g^{\mu \nu} q_{1,\mu} q_{1,\nu}$ for the corresponding geometry. $I_{Schw}, \, I_{III}$ are zero only for the geometry for which they have been calculated, thus distinguishing them.

By taking a step further besides considering single states, one could try and define a generalized wave function containing all possible geometries within a certain geometrical class (GC) as a sum of all different states $h(q)$,
\be \label{psifull}
\Psi_{GC} = \int\!\frac{(-h[q])^{\frac{1}{4}}}{\sqrt{\mu[q]}}\, e^{\ima\, \kappa \int\!\!\sqrt{-h[q]}\, d q}\, \mathcal{D}h(q)\mathcal{D}\delta(I(q)),
\ee
where $I(q)$ are the appropriate invariant relations corresponding to a specific base manifold geometry. It should be noted that this expression is symbolic, in the sense that it must be considered as a sum over continuous functions not as an actual integration. As far as the state $h$ is concerned, it is meant to describe different configurations of the base manifold through the particular form of all $f^i (q)$. Each set of the latter (for a given $h$) corresponds to a different geometry of the base manifold, since - in the context of the reduced system - the $f^i (q)$ need not satisfy the original classical equations.
%-------------------------------------------------------------------------------------------------------------------------------%-------------------------------------------------------------------------------------------------------------------------------
%-------------------------------------------------------------------------------------------------------------------------------%-------------------------------------------------------------------------------------------------------------------------------
\section{Discussion}
%-------------------------------------------------------------------------------------------------------------------------------
The quantization of pure gravity has turned out to pose difficult enough obstacles,
the most notable being the problem of time \cite{Kuchar:1991qf,Isham:1992ms}. This
problem appears explicitly in the canonical procedure for the quantization and has many manifestations, some of which are the absence of a unique choice of an evolution parameter, resulting in the non existence of a natural Hilbert space etc. Usually, there are two approaches of canonical quantization to
follow: (i) one first classically restricts the dynamics on the physical phase
space, by selecting gauge conditions, thus making the first class constraints second class; then performs the quantization, leading to a Schr\"odinger type equation. (ii) One performs the quantization on the full phase space.
The latter case leads to the Wheeler-DeWitt equation, the solution of which still
remains the holy grail of quantum gravity. An important issue to consider is the
consistency and preservation of the spacetime covariance of the theory after
canonical quantization. The problem is somewhat easier to be addressed in the context of the minisuperspace
models, where the diffeomorphism invariance of the theory has been
shrunk to the reparametrization invariance of the independent variable. In \cite{Christodoulakis:2004nw}, an answer is given by imposing on the wave function, apart from the quadratic constraint, the quantum analogues of the spacetime invariant relations; at the classical level, these relations are weakly vanishing quantities and thus their quantum analogues also annihilate the wave function (if correctly factor-ordered). In \cite{Barvinsky:2013aya}, the demand for consistency of the
Dirac quantization, after a procedure that takes account of the physical wave function defined on the reduced phase space, results in the Hermiticity of the momentum operator. Finally, in \cite{Salisbury:2015goa} the authors reconsider Kuchar's
proposal of the milti-fingered time, in order to regain the diffeomorphism invariance
of the Hamiltonian formalism through the construction of a diffeomorphism covariant
Hamilton-Jacobi equation. Upon quantization, it turns out that the Wheeler-DeWitt
equation is equivalent to an intrinsic Schr\"odinger equation that results from a
particular choice of an intrinsic evolution parameter.

The main characteristic of the current work is the exhibition of a general methodology concerning the classical and quantum treatment of spacetimes affording description in terms of a minisuperspace action principle. The aim is to include the entire family of geometries contained in a given base manifold's line element, not just the configurations satisfying the classical Einstein equations of motion. On the way towards this goal, the invariance under arbitrary reparametrizations of the independent evolution variable emerges as a key instrument in a twofold way:
\begin{enumerate}[(a)]
\item In the introduction, it is clearly demonstrated how the invariance of the action under the above mentioned change of independent dynamical variable generates the freedom to arbitrarily rescale the lapse (density) by any function of the scalar degrees of freedom. Subsequently, this leads to the possibility of selecting a lapse parametrization in which the potential term does not depend on any scalar degrees of freedom; thus, a particular minisuperspace metric is selected which is ``natural'' in the sense that the Euler-Lagrange equations become pure geodesics (not driven) of this metric.
\item The desire to describe the family of geometries in a way that is independent of the particular choice of the evolution parameter leads to the expression of the $(n-1)$ scalar degrees of freedom in terms of one, as $f^i (q)$ \eqref{fis}. Further use of the constraint equation by substitution into the rest of the Euler-Lagrange equations leads to reduction of the system into the $(n-1)$ second order equations \eqref{accelerations}.
\end{enumerate}
Subsequently, if one considers the $f^i (q)$'s as arbitrary, albeit given, functions of $q$, one is able to reduce the initial system \eqref{Lag} to a configuration of only two non-dynamical degrees of freedom (the lapse and one more arbitrary scalar degree of freedom). By performing the Hamiltonian analysis, we find that the reduced system still remains constrained and in addition it has an integral of motion. During the quantization procedure, we impose the quantum version of this conserved quantity together with the Hamiltonian constraint on the wave function, thus obtaining an additional eigenvalue equation. The outcome is a solution for the wave function and an expression for the measure in which the quantum operators are Hermitians. Therefore this line of thinking, leads us to the following important results:
\begin{itemize}
\item The first, of physical value, is an expression of a generalized probability corresponding to any configuration in the designated class of geometries.
\item The second is that this probability has its extrema on the classical solution of the initial Lagrangian.
\item  Consequently, this generalized probability, interpreted as a regular action, can also describe the ensuing true dynamics of the innitial, non-reduced system; and this is so despite the fact that it has originated from quantizing the gauge part of the initial system. We thus find, at the quantum level, a justification of the adjective dynamical atributed to the classical quadratic constraint.
\end{itemize}
Further information is obtained by examining the nature of these extrema. More specifically, the characterization of the candidates for extrema is possible whenever there exists a matrix $U$ satisfying the Riccati equation \eqref{matric} and whether the extremum is maximum or minimum depends on the positive or negative definite nature of the matrix $W$. A simple demonstration of the method was given in section 7 by some concrete examples.

\begin{acknowledgments}
N. D. acknowledges financial support by FONDECYT postdoctoral grant no. 3150016.
\end{acknowledgments}

\bibliography{Geom_clas_quant-revtex}
\end{document}